\documentclass[aps,prl,twocolumn,groupedaddress,floatfix,twoside]{revtex4}

\usepackage{graphicx}
\usepackage{dcolumn} 
\usepackage{bm}

\begin{document}

\title{Detection of spatial correlations in an ultracold gas of fermions}

\author{M. Greiner}
\email[Email: ]{markus.greiner@colorado.edu} \homepage[URL:
]{http://jilawww.colorado.edu/~jin/}
\author{C. A. Regal}
\author{C. Ticknor}
\author{J. L. Bohn}
\author{D. S. Jin}
\altaffiliation[]{Quantum Physics Division, National Institute of
Standards and Technology.}

\affiliation{JILA, University of Colorado and National Institute
of Standards and Technology,\\ and Department of Physics,
University of Colorado, Boulder, CO 80309-0440, USA}

\date{\today}
\begin{abstract}
Spatial correlations are observed in an ultracold gas of fermionic
atoms close to a Fesh\-bach resonance. The correlations are
detected by inducing spin-changing rf transitions between pairs of
atoms. We observe the process in the strongly interacting regime
for attractive as well as for repulsive atom-atom interactions and
both in the regime of high and low quantum degeneracy. The
observations are compared with a two-particle model that provides
theoretical predictions for the measured rf transition rates.
\end{abstract}

\pacs{03.75.Ss, 34.50.-s)}

\maketitle


Close to a Fesh\-bach resonance
\cite{Feshbach1962a,Stwalley1976a,Tiesinga1993a} ultracold
fermionic atoms form a strongly interacting quantum gas. Recently
it has become possible to study this exotic quantum regime with
$^{40}$K and $^{6}$Li atoms. Mean-field interactions
\cite{Regal2003b,Gupta2003a,Bourdel2003a} and hydrodynamic
behavior \cite{O'Hara2002a,Regal2003b,Bourdel2003a} have been
observed. However a Feshbach resonance does more than simply alter
the interactions between atoms. The resonance occurs when the
collision energy of two free atoms coincides with that of a
molecular state in a closed channel. Fermionic atom pairs
populating this closed channel can be described by a
composite-boson field. It has been proposed that this
composite-boson field can lead to fermionic superfluidity at
critical temperatures $T_c$ comparable to the Fermi temperature
$T_F$ \cite{Holland2001a,Holland2001b,Timmermans2001a}.

For a magnetic-field detuning on the side of the resonance with
repulsive atom-atom interactions, coupling between the open and
closed channels gives rise to a new molecular bound state. By
adiabatically scanning over a magnetic-field Fesh\-bach resonance
large numbers of these extremely weakly bound molecules have been
created reversibly in a $^{40}$K fermionic quantum gas
\cite{Regal2003a} and recently also with other fermionic
\cite{Cubizolles2003a,Strecker2003a} and bosonic
\cite{Durr2003a,Grimm2003} atomic species. Bose-Einstein
condensation of these molecules is being pursued. By inducing
radiofrequency (rf) transitions, these molecules have been
photodissociated into free pairs of atoms, and the corresponding
dissociation spectra provided precise information about molecular
wave functions and binding energies \cite{Regal2003a}.

With the opposite sign of the magnetic-field detuning this new
molecular state does not exist. However, coupling to the resonance
state in the closed channel still significantly changes the atom
pair wave function. For example, a well-studied effect of the
resonance, for either sign of the magnetic-field detuning, is to
modify the scattering phase shift at large internuclear separation
R, and thus change the scattering length. In this Letter we
instead probe the effect of a Feshbach resonance on the wave
function for an interacting atom pair at small internuclear
separation. Thereby we probe the population of the closed channel
or composite-boson field.

Coupling to the resonance state in the closed channel enhances the
amplitude of the wave function at very small internuclear
separation. The spatial size of the closed channel resonance
state, which is approximately $50\,a_0$ where $a_0$ is the Bohr
radius, is two orders of magnitude smaller than the average
separation between particles for a typical trapped gas. Thus, the
effect of the resonance is to increase the pair correlation
function at $R\sim 0$. To demonstrate how large this effect can
be, we calculate that near a Feshbach resonance the fraction of
atom pairs at $R\leq50\,a_0$ is five orders of magnitude larger
than away from the resonance.

\begin{figure}
\includegraphics[width=\linewidth]{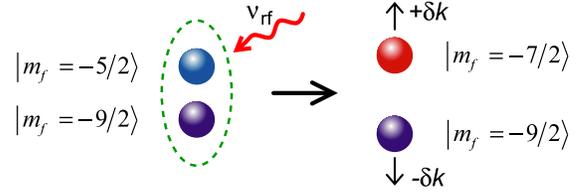}
\caption{Two-particle model of spin-changing rf transitions
between interacting pairs of atoms in the $m_f$=$-5/2$ and
$m_f$=$-9/2$ spin state. After the transition the spin state of
the first atom is changed to $m_f$=$-7/2$. In addition the
relative momentum of the atoms at large internuclear separation
can be changed by $2\delta k$ in this process if the applied radio
frequency $\nu_\mathrm{rf}$ is detuned with respect to the
$m_f$=$-5/2$ $\rightarrow$ $m_f$=$-9/2$ transition of bare atoms.
The total kinetic energy of the free atoms is changed by an amount
equal to the excess energy of the rf photon.}
\end{figure}

We probe these atom pair correlations at $R\sim 0$ by inducing a
class of spin-changing rf transitions that is only expected to
occur for spatially correlated and interacting pairs of atoms. For
bare atoms spin-changing transitions can be induced by applying an
rf field with a frequency $\nu_0$ corresponding to the Zeeman
splitting of the spin states. The transition probability
diminishes if the detuning $\Delta$ of the rf field is large
compared to the Rabi frequency $\Omega$. In contrast, an
interacting pair of atoms at small $R$ can have a significant
transition probability for relatively large detunings. In Fig.~1
we illustrate this process with a two-particle model. Here an rf
transition can occur, in which not only the spin, but also the
relative momentum of the free atom pair is changed (Fig.~1).

In the following we show that we experimentally observe rf
transitions between correlated pairs of atoms when a detuned rf
field is applied to an ultracold gas of fermions. This process is
only detected close to a Fesh\-bach resonance and thereby directly
demonstrates strong spatial correlations in this exotic quantum
regime. Although many-body physics is likely to play a significant
role in this regime, we compare the observations with a
two-particle model as a first step toward understanding the
correlations.

The experimental set-up and procedure are similar to that in our
previous experimental work
\cite{Loftus2002a,Regal2003b,Regal2003a}. In brief, we
evaporatively cool a spin mixture of $^{40}$K atoms in the lowest
hyperfine ground state. The first cooling step is carried out in a
magnetic trap. Then, the atoms are loaded into an optical dipole
trap, where further evaporative cooling is performed by lowering
the depth of the trapping potential. Finally we prepare about
10$^6$ atoms at an adjustable temperature, which for these
experiments is between $T$\,=\,0.33\,$T_F$ and 1.0$\,T_F$. The
final radial trapping frequency ranges between $\nu_r$\,=\,200\,Hz
and $\nu_r$\,=\,450\,Hz, and the axial trapping frequency is given
by the fixed ratio $\nu_r/\nu_z\!=\!70$. In the experiments the
Fermi momentum at temperatures $T$\,=\,0.33\,$T_F$ is on the order
of $k_F\!=\!(2000\,a_0)^{-1}$.

We can widely vary the interaction between atoms using an $s$-wave
Fesh\-bach resonance, which is located at a magnetic field of
$B\!=\!224.21\,\pm0.05$\,G and has a width of $9.7\pm0.6$\,G
\cite{Regal2003b}. Feshbach resonances have been exploited for
quantum degenerate Bose and Fermi gases
\cite{Inouye1998a,Cornish2000a,Loftus2002a,Dieckmann2002a,
O'Hara2002a,Regal2003c,Regal2003b,Gupta2003a,Bourdel2003a,
Regal2003a,Durr2003a,Cubizolles2003a,Strecker2003a,Grimm2003}. The
$s$-wave resonance used in this experiment affects collisions
between atoms in the $|f$=9/2,$m_f$=-5/2$\rangle$ and
$|f$=9/2,$m_f$=-9/2$\rangle$ spin states. Here $f$ denotes the
total angular momentum and $m_f$ the magnetic quantum number. The
effective scattering length of the atoms can then be varied by
tuning the strength of a homogenous magnetic field around the
Fesh\-bach resonance value.

Initially the atoms are prepared in the $f$=9/2 hyperfine state in
an incoherent spin mixture of the $m_f$=$-7/2$ and the
$m_f$=$-9/2$ states. In these states, the atoms are not affected
by the Fesh\-bach resonance. Then, after ramping to the final
magnetic-field value $B$, the atoms in the $m_f$=$-7/2$ state are
transferred into the $m_f$=$-5/2$ state by applying an rf
$\pi$-pulse between the Zeeman sublevels. In these states the
atoms are strongly interacting if the magnetic field is chosen
close to resonance. This sequence, which may result in a
nonequilibrium sample, was chosen to suppress the population of
the molecular bound state close to threshold that exists on the
repulsive side of the resonance. This is in contrast to our
previous work, where large numbers of molecules were reversibly
created by ramping the magnetic field across the Fesh\-bach
resonance.

The basic idea of the experiment is to use the rf spectroscopy
method described above to look for an enhancement in the $R\sim 0$
pair correlation function near a Feshbach resonance. We apply an
rf field pulse to induce transitions from the $m_f$=$-5/2$ to the
$m_f$=$-7/2$ state. The rf field is far detuned with respect to
the bare atom transition, with a detuning larger than the resonant
Rabi frequency $\Delta >\Omega$. Therefore, only a small fraction
of bare atoms undergo an off-resonant rf transition. The detuning
is also large compared to mean-field shifts \cite{Regal2003b}.
Nevertheless we find a significant, magnetic-field dependent
transition rate that peaks in the strongly interacting regime
close to the Fesh\-bach resonance.

In Fig.~2 the fraction of atoms transferred into the $m_f$=$-7/2$
state is plotted as a function of magnetic field $B$. We observe
pair transitions on both the repulsive (low $B$) and attractive
(high $B$) sides of the resonance. Figure~3 demonstrates that for
longer pulses the fraction of transferred atoms saturates at a
finite value.

\begin{figure}
\includegraphics[width=\linewidth]{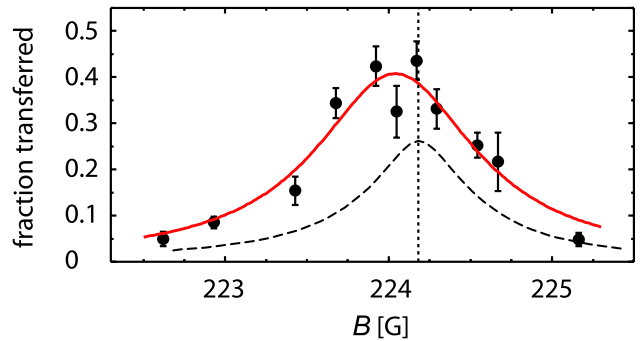}
\caption{Atom pair signal for different magnetic-field values $B$
near the Feshbach resonance. The fraction of atoms transferred
from the $m_f$=$-5/2$ into the $m_f$=$-7/2$ spin state is measured
by spin-selective time-of-flight absorption imaging after an rf
pulse with a Rabi frequency of $\Omega$=2$\pi\cdot32$\,kHz is
applied for 100\,$\mu$s. For each B, $\nu_{r\!f}$ is chosen so
that the rf field is kept at a constant large detuning
$\Delta$=$-2\pi\cdot100$\,kHz with respect to the bare atom
transition. A calculated 6.3\% of off-resonantly transferred bare
atoms has been subtracted. The solid line is a Lorenzian fit with
an amplitude of $0.41\pm0.03$ and a width of $1.2\pm0.2\,$G. It is
shifted to the repulsive side of the resonance (dotted line) by
$0.13\pm0.04$\,G. The temperature of the atoms in this measurement
was $T\!=\!0.33\pm0.06\,T_F$ and the peak density was
$n_p\!=\!(1.2\pm0.6)\times10^{13}\,$cm$^{-3}$ per spin state. The
dashed line shows a theoretical plot for a two-body multi channel
scattering theory including the rf field. The calculated maximum
fraction of atoms is 0.25 and the width is 0.79\,G.}
\end{figure}

\begin{figure}
\includegraphics[width=\linewidth]{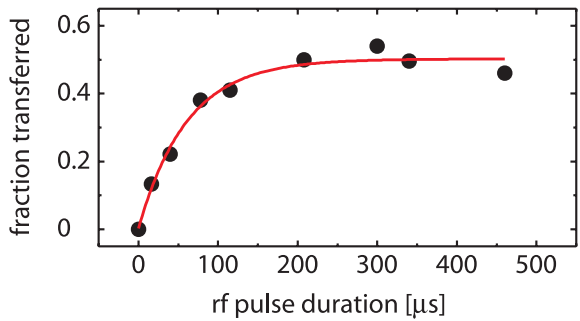}
\caption{Fraction of atoms transferred from the $m_f$=$-5/2$ into
the $m_f$=$-7/2$ spin state versus the duration of the applied rf
pulse. This measurement is performed on the attractive side of the
resonance at a magnetic field of $B\!=\!224.30$\,G, with an rf
detuning of $\Delta$=$-2\pi\cdot100$\,kHz. The solid line is a fit
to an exponentially saturating growth curve. The curve saturates
at a fraction of $0.49\pm0.03$ with a time constant of
$61\pm8\,\mu$s.}
\end{figure}

For comparison to the observed process, we have constructed a
complete two-body multichannel scattering theory that includes the
rf interaction \cite{Moerdijk1996c,Bohn1999a}. Here nearby atoms
perturb each other's internal structure so that a pair can absorb
an rf photon that is not resonant with either atom separately.
Furthermore the rf photon contributes a quantum of angular
momentum, introducing a coupling between atom pairs with
$m_{tot}=-7$ and  $m_{tot}=-8$, where $m_{tot}$ is the sum of the
spin projections of the two atoms. This two-body theory provides
transition rates for a given density, momentum distribution,
magnetic field $B$, and rf detuning $\Delta$.

The dashed line in Fig.~2 shows the results of such a calculation.
The relative momentum of the particles has been modeled by a
$T\!=\!0.34\,\mu$K Boltzmann distribution that approximates the
measured momentum distribution of the atoms. All other parameters
are identical to the measurement shown in Fig.~2. A comparison
with the calculation suggests that the experimentally observed
effect can at least in part be explained by the two-particle
model. However, the theory prediction is nearly symmetric with
respect to the resonance position whereas we measure a larger
effect that is slightly shifted to the repulsive side of the
resonance and broader. One source of this deviation may be due to
a small number of residual bound molecules that only exist on the
repulsive side and are photodissociated by the rf field. In
addition, we expect that there may be significant many-body
effects in the experiments.

An intriguing result of the calculation is that the rf pair
process occurs deep within the interatomic potential where the
spacing of the atom pair is about $22\,a_0$. At this distance an
rf-mediated spin exchange avoided crossing appears in the coupled
channel Hamiltonian. The process therefore measures spatial
correlations between atoms on a very short length scale. Close to
a Feshbach resonance these correlations are strongly enhanced by
the population of the closed channel state. Therefore the rf
process predominantly probes the composite-boson field of the
closed channel.

The rf transitions reported here are reminiscent of the recently
observed rf photodissociation of bound molecules created at a
Feshbach resonance \cite{Regal2003a}. The measurements, however,
differ from each other in several ways. In this experiment we did
not deliberately populate the molecule state. In addition, in
order to maximize the signal here we used an rf pulse duration
that is an order of magnitude longer than in the molecule
dissociation experiment. Also, within the precision of these
experiments we do not observe a frequency shift corresponding to a
binding energy. Finally the rf transitions between interacting
atom pairs are observed on both the attractive and the repulsive
sides of the Feshbach resonance. Bound molecules, on the other
hand, only exist on the repulsive side of the resonance.

In order to study the observed rf process in more detail we have
measured the kinetic energy of the transferred atoms. If the
observed rf transitions for large rf detunings can be attributed
to pairs of atoms as illustrated in Fig.~1, the kinetic energy of
each transferred atom should increase by half the excess energy of
the rf photon as \cite{Regal2003a}
\begin{equation}
\delta E_{kin}\!=\!-0.5\cdot(\,h\nu_{r\!f}-h\nu_\mathrm{0}\,).
\end{equation}
Here $\nu_\mathrm{0}$ is the rf transition frequency for bare
atoms including mean-field shifts and $\nu_{r\!f}$ is the
frequency of the rf field. The prefactor is negative since we
observe an rf transition to a lower lying Zeeman state, where an
rf photon is emitted in a stimulated process. Therefore, we expect
that the atoms gain kinetic energy when the rf is detuned to lower
frequencies.

\begin{figure}
\includegraphics[width=\linewidth]{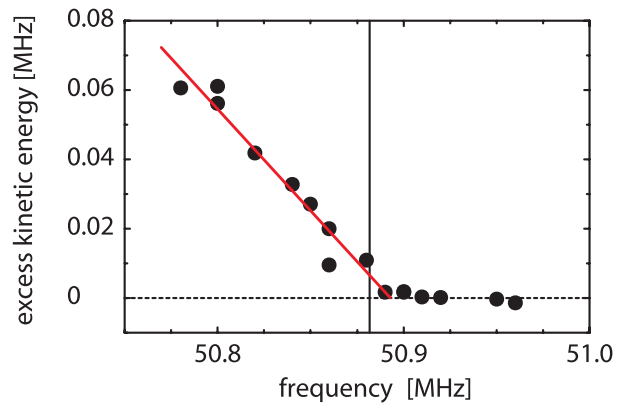}
\caption{Increase of the kinetic energy $\delta E_{kin}$ of the
transferred atoms versus the frequency $\nu_{r\!f}$ of the applied
$100\,\mu$s rf pulse, on the attractive side of the resonance. The
solid line is a linear fit to the data yielding a slope of $-0.59
\pm0.05$. This result is close to the expected value for an atom
pair process. The vertical line indicates the expected resonant rf
transition frequency for bare atoms at
$\nu_\mathrm{0}$\,=\,50.882\,MHz, neglecting mean-field shifts
that are on the order of -1 kHz. The measured zero crossing at
$\nu_{r\!f}\,=\,50.893\pm0.005$\,MHz agrees fairly well with this
value.}
\end{figure}

\begin{figure}
\includegraphics[width=\linewidth]{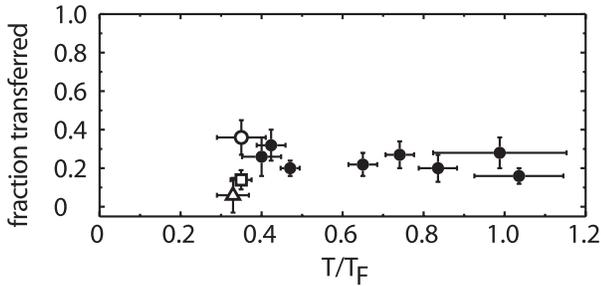}
\caption{Fraction of atoms transferred by an $100\,\mu$s rf pulse
with a detuning of $\Delta$=$-2\pi\cdot100$\,kHz for various
temperatures $T$ on the attractive side of the resonance
($B\!=\!224.37$\,G). No explicit temperature or degeneracy
dependence is observed. The data points at low temperature show a
density dependence, where $n_p=5.0\times10^{12}\,$cm$^{-3}$
(triangle), $1.2\times10^{13}\,$cm$^{-3}$ (square) and
$2.8\times10^{13}\,$cm$^{-3}$ (open circle).}
\end{figure}

Experimentally the energy of the transferred atoms is determined
by fitting time-of-flight expansion images of atoms in the
$m_f$=$-7/2$ spin state to a Gaussian model \cite{gaussfit}.
Figure~4 shows a plot of the measured increase in kinetic energy
of the atoms transferred versus the rf frequency, at
$B$\,=\,224.40\,G on the attractive side of the Feshbach
resonance. For rf frequencies lower than the resonant frequency
$\nu_{0}$, the kinetic energy indeed increases linearly with a
slope $\delta E_{kin}/(h\nu_{r\!f}-h\nu_\mathrm{0})=-0.59\pm0.05$.
The same measurement on the repulsive side of the Feshbach
resonance at $B$\,=223.84\,G yields a slope of $-0.51\pm0.06$.
This result is in reasonable agreement with the expected value in
eq. 1 and demonstrates that the excess energy of the applied rf
photon corresponds to the increase in kinetic energy of the atoms.
Due to energy and momentum conservation, this is only possible in
a process involving an interacting pair of atoms.

We find that the observed process does not strongly depend upon
quantum degeneracy. In Fig.~5 the fraction of transferred atoms on
the attractive side of the resonance is plotted for temperatures
between 0.33\,$T_F$ and 1.0\,$T_F$. We observe similar rates for
all of these temperatures and an increase of the rate for higher
densities.

In conclusion, we have observed rf transitions between strongly
interacting pairs of atoms close to a Feshbach resonance. The rf
transitions probe spatial correlations between atoms on a short
length scale. These results demonstrate that a Feshbach resonance
not only increases the strength of atom-atom interactions, but
also introduces strong pair correlations between atoms for both
positive and negative magnetic-field detunings from the resonance.
It will be interesting to study how the pairs evolve as resonance
superfluidity occurs.

\begin{acknowledgments}
It is a pleasure to thank E. A. Cornell, C. E Wieman, M. Holland,
S. Inouye and W. Zwerger for stimulating discussions and J. Smith
for experimental assistance. This work was supported by the NSF
and NIST; C.\,A.\,R. acknowledges support from the Hertz
Foundation.
\end{acknowledgments}


\begin{thebibliography}{23}
\bibitem[{gau()}]{gaussfit}
\bibinfo{note}{In order to account for a background of bare atoms that undergo
  an atomic transition due to the finite detuning and finite spectral width of
  the rf pulse, the surface fit model is a bi-modal sum of two gaussians. The
  width of the first Gaussian is fixed and measures the bare atom transition
  background, while the width of the second gaussian measures the kinetic
  energy plotted in Fig.~4.}
\end{thebibliography}

\end{document}